\documentclass[aps,prb,reprint,superscriptaddress]{revtex4-2}

\usepackage{graphicx}
\usepackage{amsmath}
\usepackage{amssymb}
\usepackage{amsthm}

\DeclareMathOperator{\var}{Var}
\def\braket#1{\langle #1\rangle}
\def\ket#1{| #1\rangle}

\begin{document}

\title{Escaping many-body localization in an exact eigenstate}

\author{Michael Iversen}
\thanks{These authors contributed equally to this work.}
\affiliation{Department of Physics and Astronomy, Aarhus University, DK-8000 Aarhus C, Denmark}

\author{N. S. Srivatsa}
\thanks{These authors contributed equally to this work.}
\affiliation{Max-Planck-Institut f\"{u}r Physik komplexer Systeme, D-01187 Dresden, Germany}
\affiliation{School of Physics and Astronomy, University of Birmingham, Birmingham, B15 2TT, UK}

\author{Anne E. B. Nielsen}
\affiliation{Department of Physics and Astronomy, Aarhus University, DK-8000 Aarhus C, Denmark}

\begin{abstract}
Isolated quantum systems typically follow the eigenstate thermalization hypothesis, but there are exceptions, such as many-body localized (MBL) systems and quantum many-body scars. Here, we present the study of a weak violation of MBL due to a special state embedded in a spectrum of MBL states. The special state is not MBL since it displays logarithmic scaling of the entanglement entropy and of the bipartite fluctuations of particle number with subsystem size. In contrast, the bulk of the spectrum becomes MBL as disorder is introduced. We establish this by studying the entropy as a function of disorder strength and by observing that the level spacing statistics undergoes a transition from Wigner-Dyson to Poisson statistics as the disorder strength is increased.
\end{abstract}

\maketitle

\section{Introduction}

Statistical mechanics is a well-established theory successfully describing quantum systems in contact with external reservoirs \cite{Kardar}. When these systems reach equilibrium, most information about the initial state is erased. In contrast, the dynamics of isolated quantum systems are determined by unitary time evolution. Recent experimental progress in preparing and controlling isolated quantum systems has drawn intense attention to the subject of describing isolated quantum systems by statistical mechanics \cite{Bloch, Blatt} as well as finding exceptions to thermal behaviors leading to interesting properties \cite{Abanin}.

The eigenstate thermalization hypothesis asserts that systems act as their own reservoir \cite{Deutsch, Srednicki}. In this way, subsystems can be in thermal equilibrium with the remaining system and expectation values of local observables then agree with those from conventional quantum statistical mechanics. While the eigenstate thermalization hypothesis makes powerful predictions about a large class of quantum systems, it is violated by various mechanisms, such as quantum integrability, many-body localization (MBL) \cite{Abanin, Nandkishore}, and quantum many-body scars \cite{Serbyn, Turner, Moudgalya}. MBL is typically achieved by introducing disorder into suitably chosen systems. This results in a complete set of quasilocal integrals of motion such that the bulk of the spectrum violates the eigenstate thermalization hypothesis. Quantum many-body scars instead provide examples of weak violations of the eigenstate thermalization hypothesis, in which non-thermal states are embedded in a spectrum of thermal states.

Symmetries play an important role in physics leading to a variety of exotic phenomena \cite{Gross}. In the context of physics of localization, the presence/absence of symmetries leads to different transport behavior and has been observed experimentally \cite{Hainaut}. Interestingly, in MBL systems, symmetries may lead to ordered eigenstate phases or even lead to breakdown of localization \cite{Parameswaran}. When a generic eigenstate at infinite temperature is invariant under a continuous non-Abelian symmetry, such as SU(2), it cannot be area-law entangled and the entanglement entropy scales at least logarithmically with system size  \cite{Protopopov}. This incompatibility between MBL and SU(2) symmetry was exploited recently to construct a model that weakly violates MBL by embedding a special eigenstate with an emergent SU(2) symmetry into an MBL spectrum \cite{Srivatsa}.

A natural question arises if the presence of a non-Abelian symmetry is the only route to avoid localization in a many-body eigenstate of a strongly disordered system. We show that this is not the case by constructing a model Hamiltonian with a special eigenstate that does not have additional symmetries compared to the Hamiltonian. The system many-body localizes when disorder is added, but the special state continues to have non-MBL properties. This suggests that partial solvability of a disordered model with suitable, exact eigenstates can provide a mechanism to achieve weak violation of MBL.

The paper is structured as follows. In Sec.\ \ref{sec:state}, we present the special state and discuss how disorder enters the model. In Sec.\ \ref{sec:disHS}, we introduce a Hamiltonian for which the special state is an exact eigenstate. In Sec.\ \ref{sec:properties}, we show that the entanglement entropy of the special state scales logarithmically with the subsystem size for both weak and strong disorder, and hence the state is neither thermal, nor MBL. Furthermore, we demonstrate that the bipartite fluctuations of particle number do not signal a transition to MBL as disorder is introduced. In Sec.\ \ref{sec:exc}, we show that the eigenstates of the considered Hamiltonian generally many-body localize by studying the entanglement entropy and the level spacing statistics. In Sec.\ \ref{sec:mid}, we investigate a modification of the model that allows us to place the special state in the middle of the spectrum, while still achieving that the remainder of the spectrum is MBL. The conclusions are summarized in Sec.\ \ref{sec:conclusion}.

\section{Special state}\label{sec:state}

We first define the special state with and without disorder. Consider a system of $N/q$ particles sitting on a lattice of $N$ sites. For odd $q$, the particles are fermions while even $q$ corresponds to hardcore bosons. The two basis states of the $j$'th site are denoted by $\ket{n_j}$ with $n_j\in\{0,1\}$. The special state is given by
\begin{multline}\label{psiLL}
\braket{n_{1}, n_{2},\ldots,n_{N}|\psi} \propto \prod_{i<j} (z_{i}-z_{j})^{qn_{i}n_{j}-n_i-n_j},
\end{multline}
in terms of $z_{j}=e^{i \phi_{j}}$ and the phases $\phi_{j} \in [0, 2\pi[$, where $j\in \mathbb Z_N = \{0, 1,\ldots,N-1\}$.

\begin{figure}	\includegraphics[width=\linewidth]{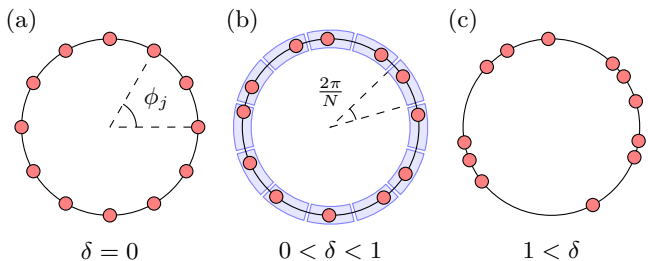}
\caption{The model is defined in terms of the phases $\phi_j$. (a) Without disorder, the phases are evenly distributed on the unit circle. (b) Disorder is introduced by randomly choosing the phases according to Eq.\ \eqref{zdisorder}. For small disorder, $0<\delta<1$, $\phi_j$ remains within $\pi/N$ of its value at no disorder (blue areas), and the ordering coincides with the non-disordered system. (c) For large disorder, $1<\delta$, the original ordering of the phases is broken and $f$ (see Eq.\ \eqref{eq:order}) is applied to recover ascending ordering.}
\label{fig:differentDisorder}
\end{figure}

We take the uniform case $\phi_j=2\pi j/N$ as a starting point and add disorder by choosing a set of random numbers $\alpha_j$ from the uniform probability distribution across the interval $[-\frac{\delta}{2}, \frac{\delta}{2}[$, where $\delta \in [0, N]$ is the disorder strength. The phases are then given by,
\begin{equation}\label{zdisorder}
\phi_{f(j)} = \phi'_j = \operatorname{mod_{2\pi}}\left( \frac{2\pi(j+\alpha_{j})}{N} \right), \quad j\in \mathbb Z_N,
\end{equation}
where the function $\operatorname{mod_{2\pi}}$ returns the remainder after division by $2\pi$ and the function $f$ orders the phases in ascending order. Let $n(A)$ denote the number of elements in a set $A$. Then $f$ is explicitly given by
\begin{align}\label{eq:order}
	\begin{split}
		f: \quad \mathbb Z_N &\to \mathbb Z_N \\
		j &\mapsto n(\{\phi'_{k}| k \in \mathbb Z_N, \phi'_{k}<\phi'_{j} \}).
	\end{split}
\end{align}
To understand the purpose of this function, it is helpful to consider the system at different disorder strengths as illustrated in Fig.\ \ref{fig:differentDisorder}.
At no disorder, $\delta=0$, we recover the uniform model and the phases are equidistant. For small non-zero disorder, $0<\delta<1$, the phases slightly differ from the phases at zero disorder. Each phase remains within $\pi/N$ of the corresponding phase at no disorder. Hence, it is not possible for two phases to get interchanged and the ordering of the phases is preserved. The function $f(j)$ is simply equal to $j$ in both of these cases. For larger disorder, $1<\delta$, the ordering of the phases may change and in the extreme case, $\delta=N$, the phases can be anywhere on the unit circle. For this case, the function $f$ relabels the phases such that the phases once again appear in ascending order. Thus, if $\phi'_j$ is the $k$'th smallest phase modulo $2\pi$ then $f(j) = k$.

\section{Hamiltonian} \label{sec:disHS}

We now construct a Hamiltonian for which $\ket{\psi}$ is an exact zero energy eigenstate. Let $d_i$ be the operator which annihilates a particle at site $i$ and $n_i = d_i^\dagger d_i$ the number operator on site $i$. Furthermore, we define the scalars,
\begin{align}
	w_{ij} = \frac{z_i + z_j}{z_i - z_j} = -i\cot \left(\frac{\phi_i - \phi_j}{2}\right).
	\label{eq:omega}
\end{align}
It was shown in \cite{Hong-Hao_Tu} that $\ket{\psi}$ is annihilated by the operators,
\begin{subequations}
	\begin{align}
		&\Lambda_{i}=(q-2)d_i+\sum_{j(\neq i)}w_{ij}[d_{j}-d_{i}(qn_{j}-1)],\\
		&\Gamma_{i}=\sum_{j(\neq i)}w_{ij}d_{i}d_{j}.
	\end{align}
	\label{operators}%
\end{subequations}
Thus, a Hamiltonian constructed as a linear combination of $\Lambda_i^\dagger \Lambda_i$ and $\Gamma_i^\dagger \Gamma_i$ has $\ket{\psi}$ as a zero energy eigenstate. Here, we consider the model described by the Hamiltonian,
\begin{equation}\label{HamiltonianInitial}
H=\sum_{i}\Lambda^{\dagger}_{i}\Lambda_{i}-q\sum_{i}\Gamma^{\dagger}_{i}\Gamma_{i}.
\end{equation}
We assume throughout the lattice filling factor is one third, i.e.\ $q = 3$.

Combining Eqs.\ \eqref{operators} and \eqref{HamiltonianInitial} yields,
\begin{equation}\label{ham}
H=\sum_{i\neq j}(F^{A}_{ij}d_{i}^{\dagger}d_{j}+F_{ij}^{B}n_{i}n_{j})+\sum_{i}F^{C}_{i}n_i+F^{D},
\end{equation}
where the coefficients are given by,
\begin{subequations}
\begin{align}
	F^{A}_{ij}&= 2\omega_{ij}(1-\omega_{ij}) - 1, \label{eq:FA}\\
	F^{B}_{ij}&= 6w_{ij}\sum_{l(\neq i, \neq j)}w_{il}, \label{eq:FB}\\
	F^{C}_{i}&= -2 \sum_{j (\neq i)} \omega_{ij}^2 - \sum_{j (\neq i)} \sum_{l (\neq i, \neq j)} \omega_{ij} \omega_{il} \label{eq:FC},\\
	F^{D}&= - \frac{N^3}{9} + N^2 - \frac{5N}{3}.\label{eq:FD}
\end{align}\label{eq:couplingCoefficients}%
\end{subequations}
The Hamiltonian conserves the number of particles. We introduce disorder to the Hamiltonian in the same way as to the wave function, namely by choosing the phases $\phi_j$ as in Eq.\ \eqref{zdisorder}.

\begin{figure}
\includegraphics[width=\linewidth]{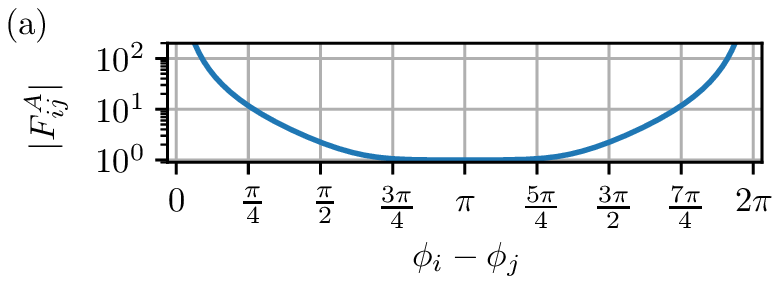}\\
\includegraphics[width=\linewidth]{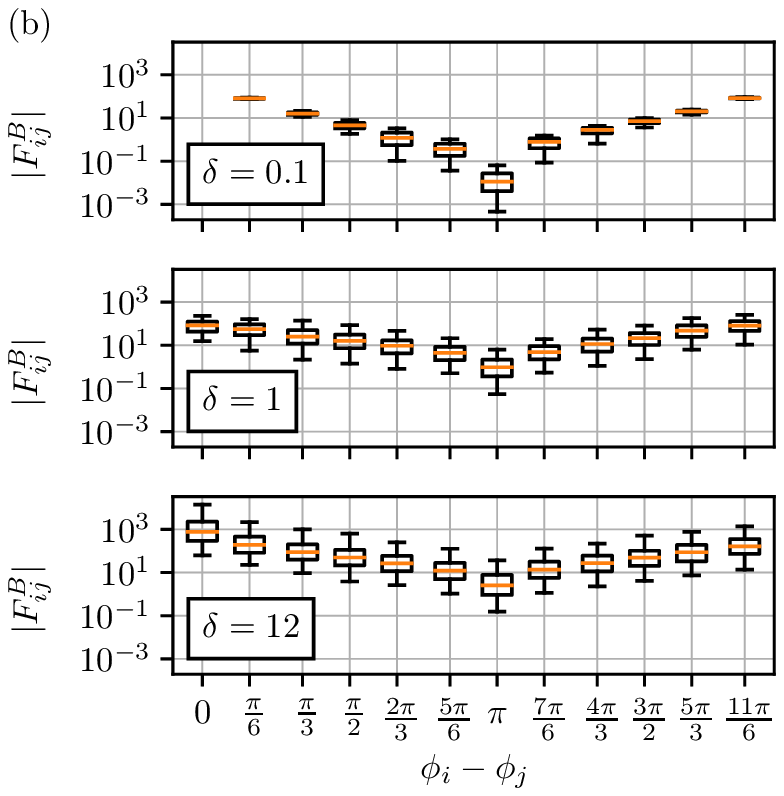}\\
\includegraphics[width=\linewidth]{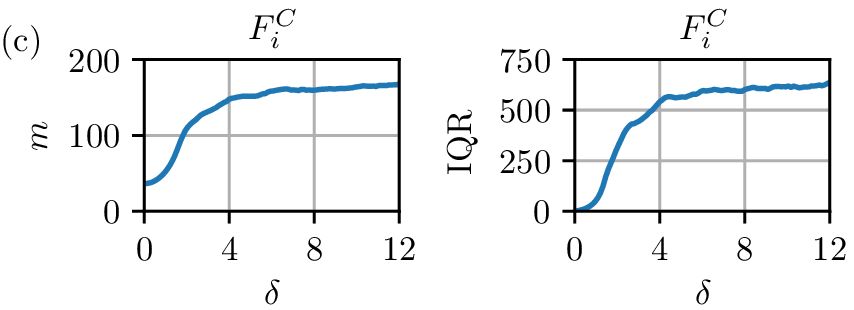}
\caption{Behavior of the coefficients in the Hamiltonian \eqref{ham}. (a) $|F^A_{ij}|$ decreases monotonically with increasing phase difference $\phi_i - \phi_j$ on the interval $[0, \pi]$ and increases on the interval $[\pi, 2\pi[$. (b) For three disorder strengths, $\delta = 0.1$, $\delta = 1$, and $\delta = 12$, the phases are constructed according to Eq.\ \eqref{zdisorder} and two sites $i$ and $j$ are chosen at random. Both $|F^B_{ij}|$ and $\phi_i-\phi_j$ are computed for $10^4$ disorder realizations and the results are grouped in $12$ intervals $\phi_i - \phi_j \in [-\frac{\pi}{12}, \frac{\pi}{12}[$, $[\frac{\pi}{12}, \frac{3\pi}{12}[$, $[\frac{3\pi}{12}, \frac{5\pi}{12}[$, etc. Each group corresponds to one box plot with the box containing $50\%$ of the data and the whiskers containing $90\%$ of the data. The median is shown as an orange line. While $|F^B_{ij}|$ varies between disorder realizations, the coefficient is typically largest when $\phi_i$ and $\phi_j$ are close together. Also note that $|F^B_{ij}|$ increases with increasing disorder strength. (c) $|F^C_i|$ is computed for $10^5$ disorder realizations and the figure illustrates the median and interquartile range (middle $50\%$ of the data). Initially, both quantities increase with disorder strength but saturates at large disorder. As discussed in the main text, both $|F^A_{ij}|$, $|F_{ij}^B|$, and $|F_{i}^C|$ diverge in the limit $\phi_i-\phi_j \to 0$. Hence the general behavior is best described by the median and interquartile range since these statistics are not sensitive to outliers (as opposed to e.g.\ the mean and variance). For all the plots $N=12$.}\label{fig:coplingCoefficients}
\end{figure}

Figure \ref{fig:coplingCoefficients} illustrates the general behavior of the coupling coefficients. $F_{ij}^A$ describes the hopping amplitude from site $j$ to site $i$ and its absolute value decreases monotonically with distance. The coefficient $F_{ij}^B$ is the interaction between particles at sites $i$ and $j$. This coefficient has a complicated behavior since the interaction strength between sites $i$ and $j$ depends on the values of all the phases. At all disorder strengths, $|F_{ij}^B|$ is typically largest when $\phi_i$ and $\phi_j$ are near each other. When increasing the disorder strength, both $|F_{ij}^B|$ and its variance between different disorder realizations generally increase. $F_i^C$ is the potential at site $i$. Finally, $F^D$ is an energy offset which ensures $\ket{\psi}$ has zero energy but does not affect the eigenstates. The coupling coefficients of hopping, interaction, and potential terms can get arbitrarily large when $\delta > 1$ since the scalars $\omega_{ij}$ diverge when $\phi_i - \phi_j \to 0$.

The purpose of this paper is to demonstrate that weak violation of MBL is possible without utilizing non-Abelian symmetries, and we construct the Hamiltonian above, because it provides a particularly clear example of this. In particular, it has the advantage that it allows us to study the special state for large system sizes with Monte Carlo techniques. Future works could use the ideas presented here to uncover models which are instead particularly suited for experimental demonstrations of weak violation of MBL without non-Abelian symmetries.

In addition to the Hamiltonian \eqref{ham}, we shall also study the Hamiltonian $H_\alpha=(H-\alpha)^2$ below, where $\alpha$ is a real number. The special state \eqref{psiLL} is an eigenstate of $H$ with eigenvalue zero, and our numerical computations show that the special state is typically either the ground state or a low-lying excited state. The Hamiltonian $H_\alpha$ has the same eigenstates as $H$, but it allows us to adjust the position of the special state in the spectrum.

\section{Properties of the special state}\label{sec:properties}

It has been shown in \cite{Hong-Hao_Tu} that important properties of the state without disorder are described well by Luttinger liquid theory with Luttinger parameter $K=1/q$. In this section, we show that the R\'enyi entropy and the bipartite fluctuation of particle number continue to scale logarithmically in the presence of disorder. This shows that the state does not many-body localize. We are capable of studying large system sizes by applying Metropolis Monte Carlo methods.

\subsection{R\'enyi entropy} \label{sec:RenyiEntropy}

The R\'enyi entropy of second order for a subsystem consisting of $L$ sites is given by
\begin{equation}
S^{(2)}_L=-\ln\left[\operatorname{Tr}\left(\rho^2_L\right)\right],
\label{eq:renyiEntropy}
\end{equation}
where $\rho_L$ is the reduced density matrix of the subsystem. We shall here take the $L$ sites to be site number $0$ to $L-1$. The R\'enyi entropy can be computed efficiently with Monte Carlo methods using the replica trick \cite{Cirac}. This is done by noting that,
\begin{equation}
\begin{split}
&\exp\left[{-S^{(2)}_L}\right]=\\
&\sum_{n,n',m,m'}|\langle n,m|\psi\rangle|^2|\langle n',m'|\psi\rangle|^2\frac{\langle\psi|n',m\rangle\langle\psi|n,m'\rangle} {\langle\psi|n,m\rangle\langle\psi|n',m'\rangle},\\
\end{split}\label{replica}
\end{equation}
where $|n\rangle$ and $|n'\rangle$ describe an orthonormal basis in the subspace of $L$ sites while $|m\rangle$ and $|m'\rangle$ describe an orthonormal basis in the subspace of the remaining $N-L$ sites. The right hand side of Eq.\ \eqref{replica} is then computed using Metropolis Monte Carlo sampling. For critical systems described by a conformal field theory, the R\'enyi entropy generally takes the form \cite{Calabrese, Vidal},
\begin{equation}
S^{(2)}_L=C\ln\left[\sin\left({\frac{\pi L}{N}}\right)\right]+\alpha,
\label{eq:criticalEntropy}
\end{equation}
where $C$ is a universal constant determined by the central charge. For the special state at zero disorder, it takes the value $C=1/4$. For Luttinger liquids, there is also a correction to the above expression that leads to $q$-periodic oscillations of the entropy \cite{calabrese2010}.

\begin{figure}
\includegraphics[width=\linewidth]{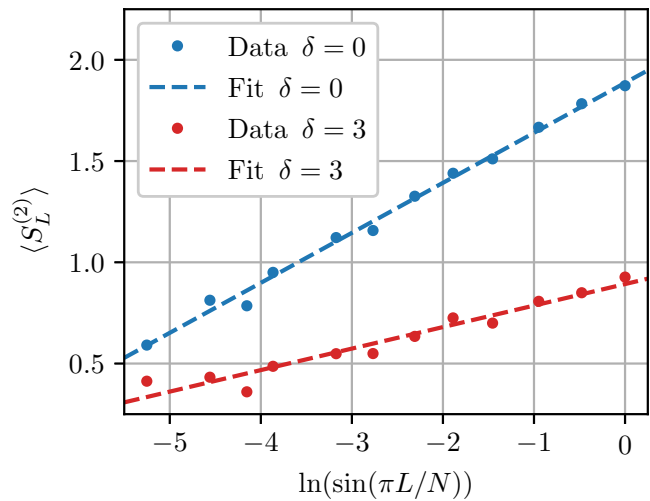}
\caption{R\'enyi entropy $S_L^{(2)}$ as a function of subsystem size $L$ for a system with total size $N=600$. The illustrated subsystem sizes $L = 1, 2, 3, 4, 8, 12, 19, 29, 45, 76, 128, 300$ respect the expected $3$-periodic deviation from Eq.\ \eqref{eq:criticalEntropy} and the corresponding values $\ln(\sin(\pi L/N))$ are approximately equidistant. For both the uniform $\delta=0$ and disordered $\delta=3$ system, the R\'enyi entropy follows Eq.\ \eqref{eq:criticalEntropy}. Without disorder, the fit $y=0.247x + 1.88$ agrees with the expected slope $C=0.25$. With disorder, $\braket{S_L^{(2)}}$ is calculated from $3000$ disorder realizations, and the fit is given by $y=0.107x + 0.893$. Since the R\'enyi entropy scales logarithmically with subsystem size, the special state $\ket \psi$ is not MBL in the presence of disorder.}\label{fig:RenyiEntropy}
\end{figure}

Figure \ref{fig:RenyiEntropy} shows the R\'enyi entropy as a function of subsystem size in the absence ($\delta = 0$) and presence ($\delta = 3$) of disorder for $N=600$. The figure also includes linear fits for both data sets. In the uniform system, $\delta = 0$, the fit is given by $y=0.247x+1.88$, and the slope agrees with the constant $C = 0.25$. For the disordered system $\delta = 3$, the data follows equation \eqref{eq:criticalEntropy} with the linear fit given by $y=0.107x + 0.893$. We obtain a similar value for the slope for a system with $N=60$ sites, and hence we do not expect the slope to change with system size. Hence, the Rényi entropy of the special state scales logarithmically with subsystem size even for strong disorder. This observation supports that the special state remains critical in the presence of disorder and conflicts with area-law scaling of entropy in MBL.

It is custom to analyze the transition to MBL using the von Neumann entropy instead of the Rényi entropy (which is also the case for our analysis in Sec.\ \ref{subsec:Entanglement_entropy}). The von Neumann entropy of a critical state described by a conformal field theory follows an expression similar to Eq.\ \eqref{eq:criticalEntropy}. Therefore, we expect the von Neumann entropy of the special state to also scale logarithmically with subsystem size. We also note that the von Neumann entropy is strictly larger than the Rényi entropy \cite{Muller-Lennert}. Consequently, when the Rényi entropy scales logarithmically, the von Neumann entropy must also scale at least logarithmically which conflict with the area-law scaling in MBL.

\subsection{Bipartite fluctuation of particle number}
\begin{figure}
\includegraphics[width=\linewidth]{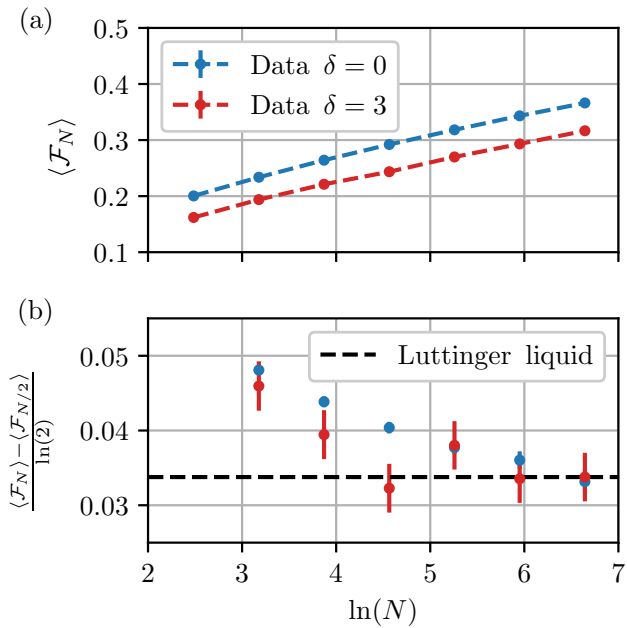}
\caption{(a) Bipartite fluctuation in particle number $\mathcal F_N$ as a function of system size $N$ without disorder $\delta = 0$ and with disorder $\delta = 3$. In both cases, the fluctuation scales logarithmically with system size. (b) The coefficient $K/\pi^2$ as a function of system size $N$. For $\delta = 0$ and $\delta = 3$, the coefficient agrees with predictions from Luttinger liquid theory $K/\pi^2 = 1/(3\pi^2)$ indicating that the state $\ket \psi$ does not many-body localize. The disorder averaged quantities are calculated by taking the mean value of $5000$ disorder realizations, and the error bars are the standard deviation of the mean over these $5000$ realizations.} \label{fig:secondCumulant}
\end{figure}

Bipartite fluctuation of particle number represents another diagnostic for identifying a transition to the MBL phase \cite{Song_2012,Luitz_2015,Singh_2016,Jean-Marie}. While the full set of even cumulants of charge fluctuations provide equivalent information to the Rényi and von Neumann entropies in some systems (e.g.\ non-interaction fermionic systems \cite{Song_2012}), the bipartite fluctuation represents a distinct quantity in our interacting model. Consider the operator $\mathcal N_N = \sum_{i=0}^{N/2-1} n_i$ which counts the number of particles in half of the chain. The subscript refers to the total system size. Then the fluctuation is given by,
\begin{align}
\mathcal F_N = \braket{\mathcal N_N^2} - \braket{\mathcal N_N}^2.
\label{eq:F}
\end{align}
The expectation value $\braket{\mathcal N_N^\ell}$ for any power $\ell$ is given by,
\begin{multline}\label{eq:N_N}
\braket{\mathcal N_N^\ell} =\\
\sum_{n_0, n_1, \ldots, n_{N-1}}\left(\sum_{i=0}^{N/2-1} n_i\right)^\ell |\braket{n_0, n_1, \ldots, n_{N-1}|\psi}|^2.
\end{multline}
Using this expression, one may compute $\braket{\mathcal N_N}$ and $\braket{\mathcal N_N^2}$ with Monte Carlo simulations. Figure \ref{fig:secondCumulant} illustrates the fluctuation as a function of system size for $\delta = 0$ and $\delta = 3$.
The fluctuation of a Luttinger liquid is asymptotically given by \cite{Song_2012},
\begin{align}\label{eq:LuttingerFluctuation}
\mathcal F_N = \frac{K}{\pi^2} \ln(N) + \text{const.}
\end{align}
While the fluctuation is generally smaller for $\delta=3$ compared to $\delta=0$, the scaling with system size in both cases agree with Eq.\ \eqref{eq:LuttingerFluctuation}. We extract the coefficient $K/\pi^2$ from the data by observing that
\begin{align}
\frac{K}{\pi^2} = \frac{\mathcal F_N - \mathcal F_{N/2}}{\ln(2)}.
\end{align}
Using the data in Fig.\ \ref{fig:secondCumulant}(a), we compute $(\mathcal F_N - \mathcal F_{N/2})/\ln(2)$ for $\delta = 0$ and $\delta = 3$. These results are illustrated in Fig.\ \ref{fig:secondCumulant}(b). Both with and without disorder, this quantity is close to the Luttinger liquid value for large system sizes. Thus, the scaling of fluctuation with system size is independent of the disorder strength. This result indicates that the special state does not undergo a transition to the MBL phase as disorder is introduced.

\section{Many-body localization}\label{sec:exc}

In this section, we investigate the properties of generic eigenstates of the considered Hamiltonian by studying the entanglement entropy and level spacing statistics. For weak disorder, the entanglement entropy displays volume-law scaling with system size while it exhibits area-law scaling at large disorder. Consistent with these findings, the level spacing statistics changes from the Wigner-Dyson distribution to the Poisson distribution as the disorder strength is increased. Both diagnostics indicate that the eigenstates generally many-body localize as disorder is introduced.

\subsection{Entanglement entropy}\label{subsec:Entanglement_entropy}

\begin{figure}
\includegraphics[width=\linewidth]{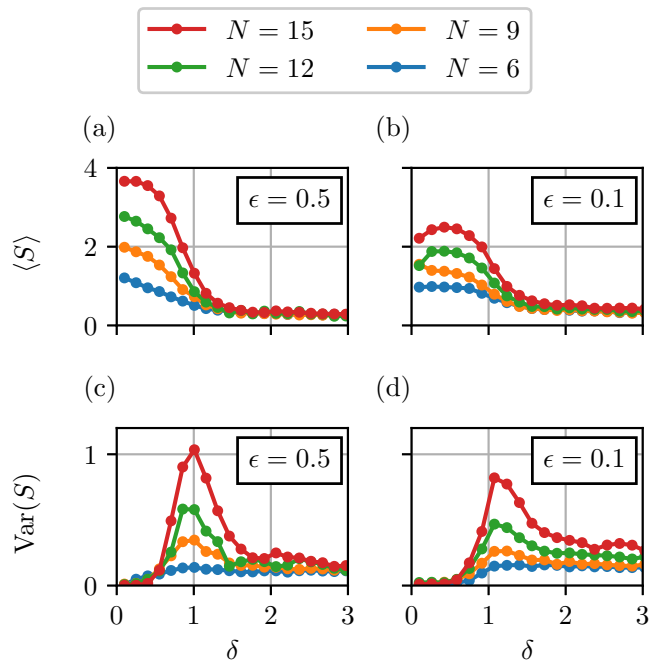}
\caption{Mean $\langle S\rangle$ and variance $\var(S)$ of the von Neumann entanglement entropy for the state closest to $\epsilon = 0.5$, i.e.\ in the middle of the spectrum, and $\epsilon = 0.1$, i.e.\ low in the spectrum, plotted against disorder strength for different system sizes $N$. Averaging is done over $10^3$ disorder realizations. At weak disorder $0<\delta \lesssim 0.5$, the system is thermal with entropy obeying volume-law scaling with system size. At intermediate disorder strengths $0.5 \lesssim \delta \lesssim 1.5$, the system transitions from thermal to MBL behavior and displays a peak in the variance of the entropy. At strong disorder $1.5 \lesssim \delta$, the system is MBL with mean entropy independent of system size.}
\label{fig:ENT}
\end{figure}

We identify the transition from thermal to MBL behavior by considering the half-chain entanglement entropy. Let $\rho = \operatorname{Tr_{\mathcal R}}(|\Psi \rangle \langle \Psi|)$ be the reduced density operator after tracing out the right half of the chain (in the case of an odd number of sites, the chain is separated into two subsystems of sizes $\lfloor N/2 \rfloor$ and $\lceil N/2 \rceil$). The von Neumann entanglement entropy is given by
\begin{align}
S = -\operatorname{Tr}[\rho \ln(\rho)].
\end{align}
The entropy of energy eigenstates follows a volume-law scaling in the thermal phase and an area-law scaling in the MBL phase. Figure \ref{fig:ENT} illustrates the mean and variance of the entropy averaged over $10^3$ disorder realizations for states at different energy densities. The energy density is defined as
\begin{equation}
\epsilon = \frac{E-E_\text{min}}{E_\text{max} - E_\text{min}},
\end{equation}
where $E_\text{min}$ ($E_\text{max}$) is the minimum (maximum) energy in the spectrum, and $E$ is the energy of the considered eigenstate. The mean and variance of the entropy are plotted as a function of disorder strength for different system sizes. These quantities display the same qualitative behavior for eigenstates in the middle of the spectrum and eigenstates low in the spectrum. For weak disorder $0 \leq \delta \lesssim 0.5$, the mean entropy scales linearly with system size consistent with the expected volume-law scaling in the thermal phase. As the disorder strength is increased, $0.5 \lesssim \delta \lesssim 1.5$, we observe a rapid increase in the variance indicating the system is undergoing a phase transition. At strong disorder, $1.5 \lesssim \delta$, the mean entropy is constant as a function of system size consistent with the area-law scaling expected in the MBL phase. These findings indicate that the system many-body localizes as disorder is introduced. This is true both in the middle of the spectrum and close to the special state low in the spectrum.

\subsection{Level spacing statistics}

\begin{figure}
\includegraphics[width=\linewidth]{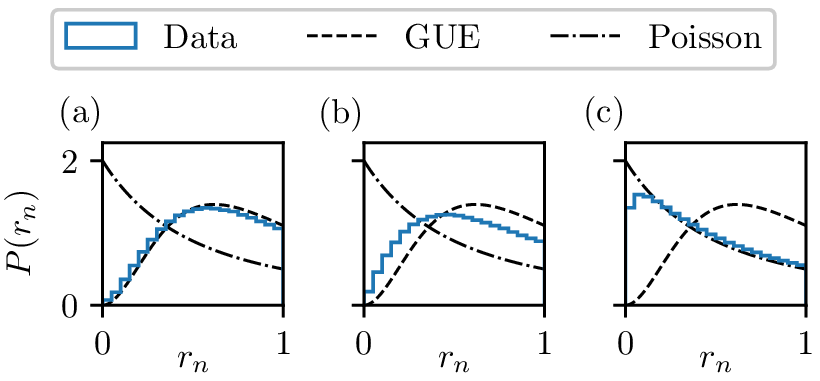}
\includegraphics[width=\linewidth]{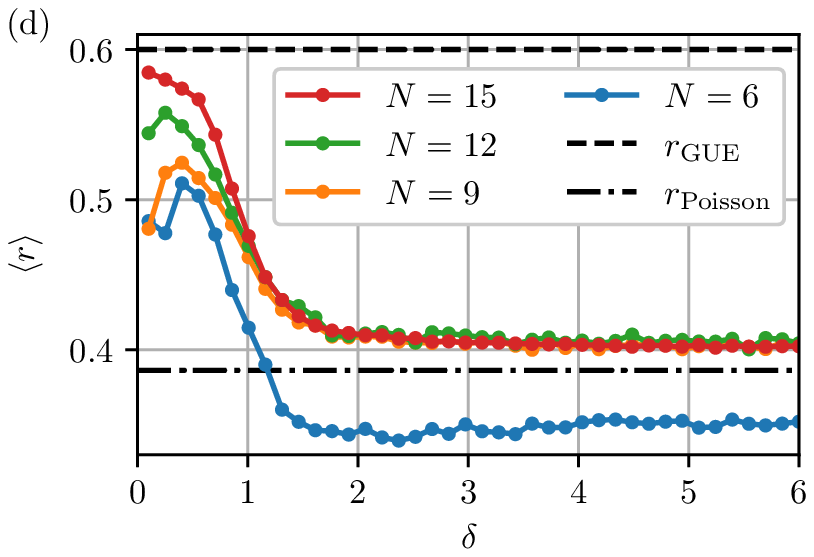}
\caption{(a)--(c) Distribution of the adjacent gap ratio obtained from $10^3$ disorder realizations and the middle third of the spectrum for system size $N=15$ at different disorder strengths (blue). For comparison, the graphs also show the corresponding GUE (dashed) and Poisson (dashed-dotted) distributions from Eq.\ \eqref{eq:GUE_Poisson}. (a) At weak disorder, $\delta = 0.1$, the distribution of the adjacent gap ratio agrees with GUE. (b) At intermediate disorder strength, $\delta = 0.75$, the adjacent gap ratio transitions from GUE towards the Poisson distribution. (c) At strong disorder, $\delta = 2$, the adjacent gap ratio agrees with the Poisson distribution. (d) The adjacent gap ratio averaged over $10^3$ disorder realizations and the middle third of the spectrum as a function of disorder strength for different system sizes. The system is thermal at low disorder and the average adjacent gap ratio agrees with GUE. As the disorder strength is increased, the adjacent gap ratio approaches the Poisson value signalling a transition to the MBL phase.}
\label{fig:gratio}
\end{figure}

The MBL phase can be identified by studying the level spacing statistics. Let $\{E_n\}$ be the energy levels sorted into assenting order and $s_n = E_{n+1}-E_{n} \geq 0$ the $n$'th level spacing. In the thermal phase, one expects level repulsion and the distribution of level spacings follows the Wigner surmise. Since the coefficients $F^A_{ij}$ in Eq.\ \eqref{eq:couplingCoefficients} are complex, the Hamiltonian \eqref{ham} is not invariant under time reversal and the level spacing is described by the Gaussian unitary ensemble (GUE). In the MBL-phase, no level repulsion is expected and the energy levels follow the Poisson process. The transition to MBL can be identified by observing the level spacing distribution shifting from GUE to Poisson as disorder is introduced. However, before comparing the level spacing distribution to either GUE or Poisson one must account for the local density of states, which is done with a technique known as unfolding \cite{Abul-Magd, Guhr}. The procedure of unfolding the spectrum is costly numerically and may introduce error. Therefore, it is custom to study the adjacent gap ratio $r_n$ instead of working directly with the level spacing distribution \cite{Oganesyan}. The adjacent gap ratio is given by
\begin{equation}\label{gapratio}
	r_n = \frac{\min(s_{n+1},s_{n})}{\max(s_{n+1},s_{n})}.
\end{equation}
This quantity is independent of the local density of states and no unfolding is needed. When the spectrum is described by respectively the Wigner surmise or a Poisson process, the adjacent gap ratio is distributed according to \cite{Atas}
\begin{subequations}\label{eq:GUE_Poisson}
\begin{align}
&P_\text{GUE}(r_n) = \frac{81 \sqrt 3}{2 \pi}\frac{(r_n + r_n^2)^2}{ (1 + r_n + r_n^2)^4}, \label{eq:GUE} \\
&P_\text{Poisson}(r_n) = \frac{2}{(1+r_n)^2} \label{eq:Poisson}.
\end{align}
\end{subequations}

Figure \ref{fig:gratio}(a)--(c) shows the distribution of the adjacent gap ratio (blue line) of a system with $N = 15$ sites at different disorder strengths (a) $\delta=0.1$, (b) $\delta = 0.75$, and (c) $\delta = 2$. The figure also displays the probability distributions in Eq.\ \eqref{eq:GUE_Poisson} for comparison (dashed and dashed-dotted lines). At low disorder, the adjacent gap ratio agrees with GUE indicating the system is thermal. As the disorder strength is increased, the distribution shifts towards the Poisson distribution. At large disorder, the adjacent gap ratio follows the Poisson distribution signalling the system is MBL.

The transition from GUE to a Poisson process is highlighted by considering the adjacent gap ratio averaged over the middle spectrum and many disorder realizations. When the spectrum is described by respectively the GUE or Poisson process, this average is given by $r_\text{GUE} = \frac{2\sqrt 3}{\pi} - \frac{1}{2} \approx 0.60$ and $r_\text{Poisson} = 2\ln(2) - 1 \approx 0.39$. The average adjacent gap ratio as a function of disorder strength is shown in Fig.\ \ref{fig:gratio}(d). We observe a transition from GUE to the Poisson process as the disorder strength is increased indicating the system becomes MBL at large disorder. The agreement with GUE at weak disorder is better for larger system sizes due to the smaller finite size effects. Similarly, the system size $N=6$ at strong disorder converges below the Poisson value due to finite size effects that reduce for larger system sizes.

\section{Placing the special state in the middle of the spectrum}\label{sec:mid}

The special state is either the ground state or one of the low-lying excited states of the parent Hamiltonian in Eq.\ \eqref{ham}. Other Hamiltonians may, however, be constructed where the special state appears higher in the spectrum. Consider the Hamiltonian $H_\alpha = (H - \alpha)^2$ obtained by shifting and squaring. Similar to the original Hamiltonian, this new Hamiltonian contains long-ranged interactions and hopping. The new Hamiltonian also contains more complicated terms such as 4-body interactions and correlated hopping. The new Hamiltonian $H_\alpha$ has the same eigenstates as $H$, but the spectrum is different since the energies are transformed as $E_n \to (E_n - \alpha)^2$. The special state is an eigenstate of $H_\alpha$ with energy $\alpha^2$, and by choosing $\alpha$ appropriately, the special state can be placed near the center of the spectrum.

\begin{figure}
\includegraphics[width=\linewidth]{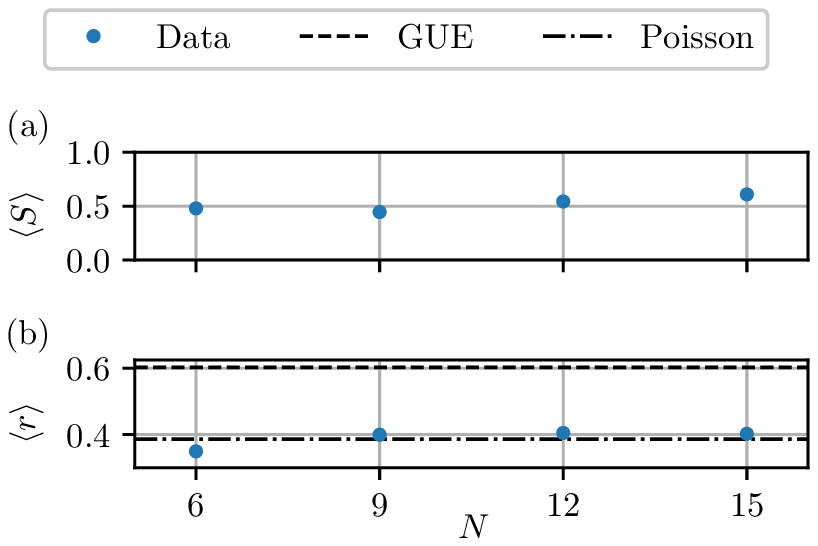}
\includegraphics[width=\linewidth]{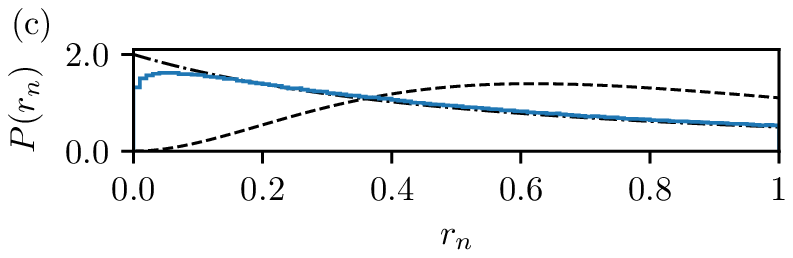}
\caption{The Hamiltonian $H_\alpha$ is investigated at strong disorder $\delta = 3$. (a) The average von Neumann entanglement entropy for the eigenstate closest to $\epsilon=0.5$, i.e.\ in the middle of the spectrum. The entropy is constant as a function of system size $N$ in accordance with area-law scaling for MBL. (b) The average adjacent gap ratio as a function of system size (blue dots), expected value for GUE (dashed line) and for the Poisson distribution (dashed-dotted line). (c) The distribution of adjacent gap ratio for system size $N=15$ (blue histogram), GUE distribution (dashed line) and Poisson distribution (dashed-dotted line). The adjacent gap ratio agrees with the Poisson distribution establishing the model is MBL. For all panels the number of disorder realizations is $10^3$, and in (b) and (c) the middle third of the spectrum has been used.}
\label{fig:shift}
\end{figure}

The new model also localizes in the presence of disorder. Figure \ref{fig:shift} illustrates the average entropy and adjacent gap ratio for $H_\alpha$ with $\alpha = E_\text{max}/(1 + \sqrt{3})$ which places the special state one third into the spectrum. We consider disorder strength $\delta = 3$ in all panels. Figure \ref{fig:shift}(a) shows that the average entanglement entropy of an eigenstate in the middle of spectrum is constant as a function of system size. In Fig.\ \ref{fig:shift}(b), we observe the average adjacent gap ratio agreeing with the expected value for the Poisson distribution. Figure \ref{fig:shift}(c) illustrates the distribution of the adjacent gap ratio for system size $N=15$ also agreeing with the Poisson distribution. These diagnostics establish the Hamiltonian $H_\alpha$ is MBL at large disorder. While the new Hamiltonian $H_\alpha$ is more complicated than the original $H$, this analysis demonstrates that the special state is not inherently limited to low energy densities. It may as well exist near the center of the spectrum. Future work could seek to uncover simpler models hosting non-MBL states in the middle of an MBL spectrum.

\section{Conclusion} \label{sec:conclusion}

While emergent symmetry has previously been identified as a mechanism to obtain weak violation of MBL \cite{Srivatsa}, we have here shown that weak violation of MBL can also happen without the presence of emergent symmetry. Specifically, we have constructed a model with a known eigenstate. Considering entanglement entropy and level spacing statistics, we have shown that the model many-body localizes at strong disorder. Nevertheless, the entanglement entropy and bipartite fluctuation of particle number for the known eigenstate scales logarithmically with system size implying that this state is not MBL at strong disorder. Our model hence contains a special non-MBL state embedded in a spectrum of MBL states. The idea to have exactly solvable eigenstates embedded in a spectrum is quite general and draws parallels to quantum many-body scars, and we expect that several further examples of weak violation of MBL can be constructed along these lines.

\begin{acknowledgments}
This work has been supported by the Carlsberg Foundation under grant number CF20-0658, the Independent Research Fund Denmark under grant number 8049-00074B, and the UKRI Future Leaders Fellowship MR/T040947/1.
\end{acknowledgments}

\end{document}